\documentclass[12pt,a4paper,twoside]{article}        
\usepackage[english]{babel}                          
\usepackage[cp1251]{inputenc}                        
\usepackage{mmgart}                                  
\usepackage{cite}

\def\ccom{\color{green!40!black}}
\def\cmath{\color{blue}}
\def\adimH{\mathsf{K}} 
\def\alert{\emph} 
\def\bmat{\begin{pmatrix}}
\def\emat{\end{pmatrix}}

\def\FBT{\mathsf{W}} 
\def\C{\mathbb{C}} 
\def\e{\mathrm{e}}

\def\Hspace{\mathcal{H}} 
\def\id{\mathbf{1}} 
\def\idmat{\mathbb{I}} 
\def\ig{\gamma} 
\def\iG{\Gamma} 
\def\iGN{{\cabs{\iG}}} 
\def\iGX{\iG^{\X}} 
 
\def\ls{\sigma} 
\def\lS{\Sigma} 
\def\lSN{{\cabs{\lS}}} 
\def\lSX{\lS^{\X}} 
\def\N{\mathbb{N}} 
\def\natmod{\mathsf{H}} 
\def\NF{\mathcal{F}} 

\def\Partransportt{\chi} 
\def\period{\mathcal{C}} 
\def\Prob{\mathrm{\mathbf{P}}} 
\def\Q{\mathbb{Q}} 
\def\R{\mathbb{R}} 
\def\regrep{\mathrm{P}} 

\def\repq{\mathrm{U}} 

\def\runisymb{\mathsf{r}} 
\def\sg{\mathsf{f}} 
\def\sG{\mathsf{F}} 
\def\sGN{{\cabs{\sG}}} 
\def\Time{\mathcal{T}} 
\def\transmatr{\mathrm{T}} 

\def\wg{\mathsf{g}} 
\def\wG{\mathsf{G}} 
\def\wGN{\mathsf{M}} 
\def\ws{\omega} 
\def\wS{\Omega} 
\def\wSN{\mathsf{N}} 
\def\x{\mathsf{x}} 
\def\X{\mathrm{X}} 
\def\XN{\mathsf{n}} 
\def\Z{\mathbb{Z}} 
\newcommand{\AltG}[1]{\mathsf{A}_{#1}} 
\newcommand{\Aut}[1]{\mathrm{Aut}\vect{#1}} 
\newcommand{\baseform}[1]{\mathcal{A}_{#1}} 
\newcommand{\barket}[1]{\left|#1\right\rangle} 
\newcommand{\brabar}[1]{\left\langle#1\right|} 
\newcommand{\braket}[1]{\left\langle#1\right\rangle} 
\newcommand{\bornform}[1]{\mathcal{B}_{#1}} 
\newcommand{\cabs}[1]{\left|#1\right|} 
\newcommand{\CyclG}[1]{\Z_{#1}} 

\newcommand{\inner}[2]{\left\langle#1\mid#2\right\rangle} 
\newcommand{\innerform}[3]{\braket{#1\cabs{#2}#3}} 
\newcommand{\IrrRep}[1]{\mathbf{#1}} 
\newcommand{\Math}[1]{$\cmath{}#1$} 
\newcommand{\Mathh}[1]{\begin{equation*}\cmath{#1}\end{equation*}} 
\newcommand{\MathEq}[1]{\begin{equation*}\cmath{#1}\end{equation*}}
\newcommand{\MathEqLab}[2]{\begin{equation}\cmath{#1}\label{#2}\end{equation}}
\newcommand{\MathEqArr}[1]{\begin{align*}\cmath{}#1\end{align*}}
\newcommand{\MathEqArrLab}[1]{\begin{align}\cmath{}#1\end{align}}
\newcommand{\Mtwo}[4]{\bmat#1&#2\\#3&#4\emat} 

\newcommand{\OrtG}[1]{\mathsf{O}\!\vect{#1,\R}} 
\newcommand{\ordset}[1]{\left[#1\right]} 
\newcommand{\Perm}[1]{\mathrm{Sym}\left(#1\right)} 
\newcommand{\PermRep}[1]{\mathbf{\underline{#1}}} 
\newcommand{\ProbBorn}[2]{\Prob\!\vect{#1,#2}} 
\newcommand{\projector}[1]{\mathrm{Pr}_{#1}}
\newcommand{\runi}[1]{\runisymb_{#1}} 
\newcommand{\set}[1]{\left\{#1\right\}} 
\newcommand{\SymG}[1]{\mathsf{S}_{#1}} 

\newcommand{\vect}[1]{\left(#1\right)} 
\newcommand{\Vtwo}[2]{\bmat#1\\#2\emat} 
\newcommand{\Vthree}[3]{\bmat#1\\#2\\#3\emat} 
\begin{document}

\setcounter{page}{1}                                
\thispagestyle{empty}                                
\begin{heading}                                      
{Volume\;3,\, N{o}\;1,\, p.\,1 -- 24\, (2015)}      
{}
\end{heading}                                        

\begin{Title}
Discrete dynamical models: combinatorics, statistics and continuum approximations
\end{Title}

\begin{center}
\Author{}{Vladimir V. Kornyak}
\end{center}


\begin{flushleft}

\Address{}{Joint Institute for Nuclear Research, Dubna, Russia}


\Email{kornyak@jinr.ru}
\end{flushleft}

\Headers{V.\,V. Kornyak}{Discrete dynamical models}

\begin{flushleft}                                 
\small\it Received 5 January 2015. Published 21 January 2015.         
\end{flushleft}                                   

\Thanks{The work is supported in part by the Ministry of Education and Science of the Russian Federation (grant 3003.2014.2) and the Russian Foundation for Basic Research (grant 13-01-00668).}

\Thanks{\mbox{}\\
\copyright\,The author(s) 2015. \ Published by Tver State University, Tver, Russia}
\renewcommand{\thefootnote}{\arabic{footnote}}
\setcounter{footnote}{0}

\Abstract{
This essay advocates the view that any problem that has a meaningful empirical content, can be formulated in constructive, more definitely, finite terms. 
We consider combinatorial models of dynamical systems and  approaches to statistical description of such models.
We demonstrate that many concepts of continuous physics --- such as continuous symmetries, the principle of least action, Lagrangians, deterministic evolution equations --- can be obtained from combinatorial structures as a result of the large number approximation.
We propose a constructive description of quantum behavior that provides, in particular, a natural explanation of appearance of complex numbers in the formalism of quantum  mechanics.
Some approaches to construction of discrete models of quantum evolution that involve gauge connections are discussed.
}



\newpage                               
\renewcommand{\baselinestretch}{1.1}   


\section{Introduction}
Any continuous physical model is empirically equivalent to a certain finite model.
This thesis is widely used in practice:  
solution of differential equations by the finite difference method or by using truncated series is typical example.
\par
It is often believed that continuous models are ``more fundamental'' than discrete or finite ones.
However, there are many indications that nature is fundamentally discrete at small (Planck) scales, and is possibly finite.%
\footnote{The total number of binary degrees of freedom in the Universe is about \Math{~ 10^{122}} as estimated via the holographic principle and the Bekenstein--Hawking formula.}
Moreover, description of physical systems by, e.g., differential equations can not be fundamental in principle, since it is based on approximations of the form
 \Math{f\vect{x}\approx{}f\vect{x_0}+\nabla{f\vect{x_0}}\Delta{}x}. 
\par
This essay advocates the view that finite models provide a more relevant description of physical reality than continuous models which are only approximations in the limit of large numbers.%
\footnote{Comparing the Planck length, \Math{\sim10^{-35}} m, with the minimum length observable in experiment, \Math{\sim10^{-15}} m, 
we may assume that the emergence of the empirically perceived continuous space is provided by averaging over about \Math{10^{20}} discrete elements.}
Using simple combinatorial models, we show how such concepts as continuous symmetries, the principle of least action, Lagrangians, deterministic evolution equations, etc. arise from combinatorial structures as a result of the large number approximation.
We also consider some approaches to the construction of discrete models of quantum behavior and related models describing evolution of gauge connections.
\par
Any statistical description assumes one or another concept of \emph{macrostate}.
We define macrostates as equivalence classes of microstates. 
This definition is especially convenient for models incorporating symmetry groups.
We distinguish two types of statistical models: 
\begin{enumerate}
	\item
\emph{Isolated system} is purely combinatorial object in the sense that ``probability'' of a microstate has \textit{a priori} nature. 
Namely, all microstates are equiprobable, so their probabilities are equal to the inverse of their total number.
The macrostates are specified by an equivalence relation on microstates.
	\item
\emph{Open system} is obtained from an isolated system by the following modification: 
The macrostates are specified by the same equivalence relation,
but the probabilities of microstates depend on some parameters, which are introduced for approximate description of interaction of the system with the environment. 
\end{enumerate}
The archetypal examples of isolated and open systems are, respectively, \emph{microcanonical} (macrostates are defined as collections of microstates with equal energies) 
and \emph{canonical} (macrostates are defined similarly, and interaction with the environment is parameterized by the temperature) \emph{ensembles}.
\par
The classical description of a (reversible) dynamical system looks schematically as follows.
There are a set \Math{X} of states and a group \Math{G_\mathrm{cl}\leq\Perm{X}} of transformations (bijections) of \Math{X}.
Evolutions of \Math{X} are described by sequences of group elements \Math{g_t\in{}G_{\mathrm{cl}}} parameterized by the \emph{continuous} time \Math{t\in\Time=\ordset{t_a,t_b}\subseteq\R}.
The observables are functions \Math{h: X\rightarrow\R}.
\par
We can ``quantize'' an arbitrary set \Math{X} by assigning numbers from a number system \Math{\NF} to the elements \Math{x\in{}X}, i.e., by interpreting \Math{X} as a basis of a module.
The quantum description of a dynamical system assumes that the module associated with the set of classical states \Math{X} is a Hilbert space \Math{\Hspace_X} over the field of complex numbers, i.e., \Math{\NF=\C}; 
the transformations \Math{g_t} and observables \Math{h}  are replaced by {unitary} \Math{U_t} and {Hermitian} \Math{H} operators on \Math{\Hspace_X}, respectively.
\par
To make the quantum description constructive, we propose the following modifications.
We assume that the set \Math{X} is finite.
Operators  \Math{U_t} belong to the group of unitary transformations of the Hilbert space \Math{\Aut{\Hspace_X}}.
Using the fact that this group contains a finitely generated --- and hence residually finite --- dense subgroup, we can replace \Math{\Aut{\Hspace_X}} by unitary representation of some finite group \Math{G} that is suitable to provide an empirically equivalent description of a particular problem.
A plausible assumption about the nature of quantum amplitudes implies that the field \Math{\C} can be replaced by an abelian number field%
\footnote{An abelian number field is an algebraic extension of \Math{\Q} with abelian Galois group.
According to the Kronecker--Weber theorem, any such extension is contained in some cyclotomic field.}
\Math{\NF}.
This field is a subfield of a certain \emph{cyclotomic} field \Math{\Q_m} which in turn is a subfield of the complex field:  \Math{\NF\leq\Q_m<\C}.
The natural number \Math{m}, called  \emph{conductor}, is determined by the structure of the group \Math{G}.
Note that the fields \Math{\Q_m} and \Math{\C} provide empirically equivalent descriptions in any applications, because \Math{\Q_m} is a dense subfield of \Math{\C} for any \Math{m\geq3}.
\par
In this paper we will assume that the time \Math{\Time} is \emph{discrete} and can be represented as a sequence of integers, typically \Math{\Time = \ordset{0,1,\ldots,T}}.
\par
Note also that the subscript \Math{a} in the notation \Math{\Hspace_a} for Hilbert spaces is overloaded and can mean, depending on context: dimension of the space, a set on which the space is spanned, a group whose representation space is \Math{\Hspace_a}, etc.

\section{Scheme of statistical description}\label{statdescr}
For convenience of presentation, let us fix some notation:
\begin{itemize}
	\item 
\Math{U} is the \alert{full set of states} of a system \Math{A}. The states from \Math{U} are usually called \alert{``microstates''} in statistical mechanics. 
	\item 
\Math{N = \cabs{U}} is the total number of microstates.
	\item
\Math{p_u} is the \emph{probability} (\emph{weight}) of a microstate \Math{u\in{U},~\sum\limits_{u\in{U}}p_u=1}.
	\item 
\Math{\sim}	is an \alert{equivalence relation} on the set \Math{U}.
	\item
Taking an element \Math{u\in{}U}, we define the \emph{macrostate} \Math{\lambda} as the equivalence class \Math{\lambda=\set{v\in{}U\mid{}v\sim{}u}}.
	\item
\Math{\Lambda} denotes the set of macrostates.
	\item
The equivalence relation \Math{\sim\,} determines  the \alert{partition} \Math{U=\coprod\limits_{\lambda\in\Lambda}\lambda.}
	\item
\Math{K=\cabs{\Lambda}} is the number of macrostates.
	\item
\Math{N_\lambda=\cabs{\lambda}} is the size of a macrostate \Math{\lambda\in\Lambda}.	
	\item
\Math{P_\lambda} denotes the probability of an arbitrary microstate from \Math{U} to belong to the macrostate \Math{\lambda}.
\end{itemize}
\paragraph{Isolated systems.}\hspace*{-9.8pt}
The probability of a microstate of an {isolated system} is defined naturally%
\footnote{This is ``the equal \textit{a priori} probability postulate'' of statistical mechanics \cite{Tolman}.} 
as \Math{p_u=1/N} for any \Math{u\in{U}}, 
and the probability of any microstate from \Math{U} to belong to the macrostate \Math{\lambda} is, respectively, \Math{P_\lambda=N_\lambda/N}. 
Since the ``probabilities'' in isolated systems have an \textit{a priori} nature, such  systems are in fact purely combinatorial objects, and we can talk about the number of combinations instead of probability.
\paragraph{Open systems}\hspace*{-5pt}interact  with the environment.
This interaction is parameterized by assigning, in accordance with some rule, probabilities to all microstates. 
That is, the probability of a microstate \Math{u\in{}\lambda} is a function 
\MathEqLab{p_u={}p_u\vect{\alpha_1, \alpha_2,\ldots}}{probparam}
of some parameters \Math{\alpha_1, \alpha_2,\ldots}. 
These parameters and function \eqref{probparam} are determined by the specifics of a particular problem.
For open systems \Math{P_\lambda=\sum_{u\in\lambda}p_u\vect{\alpha_1, \alpha_2,\ldots}}.
\paragraph{Entropy.}\hspace*{-5pt}One of the central issues of the statistical description is the search for the most probable macrostates, i.e. the macrostates with the maximum value of \Math{P_\lambda}.
Technically, entropy is defined as the logarithm of the number (or probability) of microstates that belong to a particular macrostate.
The concept of entropy is convenient for two reasons:
\begin{enumerate}
	\item 
Since the logarithm is a monotonic function, the logarithm of any 
function has the same extrema as the function itself.
	\item
If a system can be represented as a combination of two independent systems, \Math{A=A^{\prime}\otimes{}A^{\prime\prime}}, 
then the macrostates of \Math{A} can be represented as \Math{\lambda^{\prime}\otimes{}\lambda^{\prime\prime}}.
So, when computing entropy of such decompositions, we can replace multiplication by a simpler operation --- addition:
\Math{\log\vect{P_{\lambda^{\prime}}P_{\lambda^{\prime\prime}}}=\log{}P_{\lambda^{\prime}}+\log{}P_{\lambda^{\prime\prime}}}.
\end{enumerate}
\paragraph{Stirling's formula\!\!} is one of the main tools for obtaining continuum approximations of combinatorial expressions:
\MathEqLab{\ln{n!}\approx{}\underbrace{n\ln{}n-n}_{\text{superlinear}}+\underbrace{\frac{1}{2}\ln\vect{2\pi{}n}}_{\text{logarithmic}}+\underbrace{\frac{1}{12n}-\frac{1}{360n^3}+\cdots}_{\text{decreasing terms}}}{Stirlinglog} 
For our purposes it is sufficient to retain only the terms which grow with \Math{n}, i.e. the \alert{superlinear} and \alert{logarithmic} terms.
\par
\subsection{Examples of isolated and open systems}\label{examples}
To illustrate the above, let us give a few examples of isolated and open systems.
\paragraph
{Sequences of symbols. Isolated system.}
Let \Math{\Sigma} be an alphabet of size \Math{M}:
\MathEqLab{\Sigma=\vect{\sigma_1,\ldots,\sigma_m,\ldots,\sigma_M}.}{alphabet} 
The microstates are sequences of the length \Math{T} of symbols from \Math{\Sigma}:
\MathEq{u=a_1\cdots{}a_t\cdots{}a_T\in{}U,~~a_t\in\Sigma.}
The total number of microstates is \Math{N=M^T}.
Let \Math{k^u=\vect{k_1^u,\ldots,k_m^u,\ldots,k_M^u}} be the vector of multiplicities of symbols \Math{\sigma_m} in a microstate \Math{u}.
It is obvious that \Math{k_1^u+\cdots+k_M^u=T}.
We define the equivalence relation \Math{\sim} as follows
\MathEqLab{{u\sim{}v} \Longleftrightarrow {k^u=k^v\equiv{}k}.}{equivsymb}
The macrostate \Math{\lambda_k}, defined by equivalence \eqref{equivsymb}, consists of all sequences with the multiplicity vector \Math{k}.
The total number of macrostates is
\MathEq{\displaystyle
{}K=\binom{T+M-1}{M-1}\,.}
The size of the macrostate \Math{\lambda_k} is 
\MathEq{\displaystyle{}
N_{\lambda_k}=\frac{T!}{k_1!k_2!\cdots{}k_M!}\,.}
\par
Introducing the vector of ``frequencies'' \Math{\vect{f_1=k_1/T,\ldots,f_M=k_M/T}} and
applying the leading part of Stirling approximation \Math{\log{}n!\approx{}n\log{}n-n} to the entropy \Math{S_{\lambda_k}=\log{}N_{\lambda_k}} of the macrostate \Math{\lambda_k} we obtain
\Math{S_{\lambda_k}\approx{}TH\!\vect{X},} where 
\Mathh{H\!\vect{X}=-\sum\limits_{i=1}^Mf_i\log{f_i}}
is the \alert{Shannon entropy} \cite{hdcm} of a random variable \Math{X} whose \Math{M} outcomes  have probabilities \Math{f_1,\ldots,f_M}.
\subparagraph{The model of symmetric random walk}\hspace*{-12pt}\cite{Feller} is a slight modification of the above isolated system.
Alphabet \eqref{alphabet} contains now an even number of symbols \Math{M=2d}, and it is divided into two parts \Math{\Sigma=\Sigma_+\coprod\Sigma_-}, where
\Mathh{\Sigma_+=\set{\sigma_i\in\Sigma\mid1\leq{}i\leq{}d}  \text{\color{black}~~and~~}  \Sigma_-=\set{\sigma_i\in\Sigma\mid{}d+1\leq{}i\leq{}M}.}
The elements of \Math{\Sigma_+} and \Math{\Sigma_-} can be interpreted, respectively, as ``positive'' and ``negative'' unit steps in the \Math{d} directions of the integer lattice \Math{\Z^d}.
\par
The vector of multiplicities of symbols from \Math{\Sigma} can be written as \Math{k=\vect{k_+,k_-}}, where \Math{k_+=\vect{k_1,\ldots,k_d}} and  \Math{k_-=\vect{k_{d+1},\ldots,k_{2d}}}.
The equivalence of microstates \Math{u} and \Math{v} is defined now as follows
\MathEqLab{{u\sim{}v} \Longleftrightarrow {k_+^u-k_-^u=k_+^v-k_-^v}.}{symmwalk}
The partition defined by \eqref{symmwalk} is a \textit{coarsening}%
\footnote{The partition \Math{\mathcal{P}_2=\set{B_i}} of a set \Math{S} is a \emph{coarsening} of the partition \Math{\mathcal{P}_1=\set{A_j}} of the same set 
if for every subset \Math{A_j\in\mathcal{P}_1} there is a subset \Math{B_i\in\mathcal{P}_2} such that \Math{A_j\subseteq{}B_i}. 
The opposite relation among partitions is called the \emph{refinement} \cite{hdcm}.}
of the partition defined by \eqref{equivsymb}.
The numbers of equivalence classes are \emph{figurate numbers} of some \Math{d}-dimensional regular convex polytopes --- \Math{d}-dimensional analogues of the octahedron.
For example, in the case \Math{d=3} the number of macrostates is equal to the  \Math{\vect{T+1}}th \emph{octahedral number}:
\Mathh{\displaystyle{}K=\vect{T+1}\frac{2\vect{T+1}^2+1}{3}.}
\paragraph
{Multinomial distribution. Open system.}
The microstates are also sequences of symbols from alphabet \eqref{alphabet}. 
But now the presence of an environment is assumed.
The influence of the environment is parameterized by the assumption that any symbol \Math{\sigma_m\in\Sigma} comes with a fixed individual probability \Math{\alpha_m}, such that \Math{\sum\limits_{m=1}^M\alpha_m=1.}
Thus, \Math{\alpha_1^{k_1}\alpha_2^{k_2}\cdots\alpha_M^{k_M}\equiv{p_{\lambda_k}}} is the probability of a microstate from a macrostate \Math{\lambda_k}, defined by equivalence relation \eqref{equivsymb}.
The probability of a microstate  from \Math{U} to belong to the macrostate \Math{\lambda_k} is described by the \emph{multinomial distribution}:
\MathEqLab{P_{\lambda_k}=N_{\lambda_k}p_{\lambda_k}=\frac{T!}{k_1!k_2!\cdots{}k_M!}\alpha_1^{k_1}\alpha_2^{k_2}\cdots\alpha_M^{k_M}.}{multinom}
\paragraph
{Microcanonical ensemble. Isolated system.}
The concept of a microcanonical ensemble is based on the classification of microstates by energy.
More specifically, if there is a real-valued function on microstates \Math{E: {}U\rightarrow\R}, then we can impose the equivalence relation on \Math{U}:
\MathEqLab{{u\sim{}v} \Longleftrightarrow {E\vect{u}=E\vect{v}}\equiv\mathrm{E},~~u,v\in{}U.}{microcanonequi} 
Assuming that the \emph{energy} \Math{\mathrm{E}} takes a finite number of values: \Math{\mathrm{E}\in\set{\mathrm{E}_1,\mathrm{E}_2,\ldots,\mathrm{E}_K},}
we define the \emph{microcanonical ensemble} as the macrostate \Math{\lambda_k} which is an equivalence class of relation \eqref{microcanonequi}, i.e., the set of microstates with the energy \Math{\mathrm{E}_k}. More formally:
\MathEqLab{\lambda_k=\set{u\mid{}u\in{U} \wedge{} E\vect{u}=\mathrm{E}_k}~.}{microcanon}
In statistical mechanics \cite{Chandler} the microcanonical ensemble is defined in terms of the \emph{Boltzmann entropy} formula
{\MathEq{S_{\lambda_k}=\mathrm{k}_B\ln{N_{\lambda_k}}}}
or, equivalently, via the \emph{microcanonical partition function}
{\MathEq{N_{\lambda_k}=\e^{S_{\lambda_k}/\mathrm{k}_B},}}
where \Math{N_{\lambda_k}=\cabs{\lambda_k}} and \Math{\mathrm{k}_B} is the Boltzmann constant (we may assume that \Math{\mathrm{k}_B=1}). 
\paragraph
{Canonical ensemble. Open system.}
A \emph{canonical ensemble} is an open counterpart of the microcanonical ensemble. 
Macrostates of the canonical ensemble are defined by \eqref{microcanon}.
Probability \eqref{probparam}, that parameterizes the interaction with the environment in our general scheme, is now a function of a single parameter \Math{\alpha=T}, the \emph{temperature} of the  environment.
Namely, the probability of a microstate \Math{u\in{}\lambda_k} is given by the \emph{Gibbs formula}
{\MathEq{p_{\lambda_k}\vect{T}=\frac{1}{Z}{\e}^{-\mathrm{E}_k/\mathrm{k}_BT},}}
where the normalization constant 
{\MathEq{Z=\sum\limits_{k=1}^{K}N_{\lambda_k}\e^{-\mathrm{E}_k/\mathrm{k}_BT}}}
is called the \emph{canonical partition function}.
\section{Continuum approximations}\label{contappr}
In this section we show that concepts such as continuous symmetry and the principle of least action may be obtained from combinatorial models as a result of the transition to the limit of large numbers.
As an illustration, consider the \emph{open} system described by distribution \eqref{multinom}.
In the case \Math{M=2} (\emph{binomial distribution}), all calculations can be done explicitly. 
In this case both equivalence relations \eqref{equivsymb} and \eqref{symmwalk} coincide in virtue of the equality \Math{k_1+k_2=T.} 
Entropy of \eqref{multinom} for \Math{M=2} takes the form
\MathEqLab{S=\ln{}T!-\ln{}k_1!-\ln\vect{T-k_1}!+k_1\ln\alpha_1+\vect{T-k_1}\ln{\alpha_2}.}{entropyS}
Applying the growing terms of Stirling's approximation \eqref{Stirlinglog} to this formula we have \Math{S\approx{}{S_{\text{Stirling}}}=S_{\text{superlin}}+S_{\text{log}}}, where
\MathEqLab{S_{\text{superlin}}=T\ln{}T-k_1\ln\vect{\frac{k_1}{\alpha_1}}-\vect{T-k_1}\ln\vect{\frac{T-k_1}{\alpha_2}}}{slin}
and
\MathEq{S_{\text{log}}=\frac{1}{2}\ln\vect{\frac{T}{2\pi{k_1\vect{T-k_1}}}}}
are superlinear and logarithmic parts of \Math{{S_{\text{Stirling}}}}, respectively.
Since \Math{S_{\text{log}}} and its derivatives are small for large \Math{T} and \Math{k_1}, the maximum of entropy \eqref{entropyS} is close to that of its leading part \eqref{slin}.
Thus, the approximate extremum point \Math{{k_1^*}=\alpha_1T} is obtained by solving the equation 
\Mathh{\frac{\partial{}S_{\text{superlin}}}{\partial{}k_1}=-\ln\vect{\frac{k_1}{\alpha_1}}+ln\vect{\frac{T-k_1}{\alpha_2}}=0\Longleftrightarrow\frac{k_1}{T-k_1}=\frac{\alpha_1}{1-\alpha_1}.}
Further, we can expand \Math{{S_{\text{Stirling}}}} around the point \Math{{k_1^*}}.
Retaining terms up to the second order, we obtain the chain of approximations
\MathEq{S\approx{}{S_{Stirling}}\approx{}{S^*}=\left.{S_{log}}\right|_{k_1=\alpha_1T}+\left.\frac{1}{2}{\frac{\partial^2S_{superlin}}{\partial{k_1^2}}}\right|_{k_1=\alpha_1T}\times\vect{k_1-\alpha_1T}^2,}
which leads to the final formula
\MathEqLab{S\approx{}{S^*}=\ln\sqrt{{\frac{1}{2\pi{}T\alpha_1\alpha_2}}}-\frac{1}{2T\alpha_1\alpha_2}\vect{k_1-\alpha_1T}^2.}{sapprox}
\subsection{On the origin of continuous symmetries}\label{contsymm} 
It is well known \cite{Feller} that the large numbers asymptotic of the probability distribution of symmetric random walk on the lattice \Math{\Z^d} is the fundamental solution of the heat equation, also called the \emph{heat kernel}
\MathEqLab{K\vect{t,\vec{x}}=\frac{1}{\vect{4\pi{}t}^{d/2}}\exp\vect{-\frac{x_1^2+x_2^2+\cdots+x_d^2}{4t}}.}{heatkernel}
Here \Math{t\in\R_+} and \Math{x_i\in\R} are the continual substitutes for \Math{T\in\N} and for the difference \Math{k_i-k_{d+i}\in\Z}, respectively.
\par
This gives an example of the emergence of continuous symmetries from the large numbers approximation.
The symmetry group of the integer lattice \Math{\Z^d} has the structure of the semidirect product \Math{\Z^d\rtimes{}G_d}.
For simplicity, we can drop the normal subgroup \Math{\Z^d}, the subgroup of translations, as inessential for our purposes.
The group \Math{G_d} is isomorphic to the semidirect product \Math{\vect{\CyclG{2}}^d\rtimes\SymG{d}} or, equivalently, to the wreath product \Math{\CyclG{2}\wr\SymG{d}.} The size of \Math{G_d} is equal to \Math{2^dd!}.
For example, for the square lattice the group \Math{G_{d=2}} is the symmetry group of a square --- dihedral group of order \Math{8}. 
On the other hand, approximate expression \eqref{heatkernel} is symmetric with respect to the \textbf{orthogonal group} \Math{\OrtG{d}} with cardinality of continuum.
\par
The \textbf{Lorentz symmetries} --- at least in \Math{1 + 1} dimensions --- can also be obtained in a similar way. 
Let us consider approximation \eqref{sapprox} for the entropy of binomial distribution.
We introduce the following continuous substitutes
\MathEqArrLab{
\cmath{}x&\cmath:=k_1-k_2,\nonumber\\
\cmath{}t&\cmath:=T,\nonumber\\
\cmath{}v&\cmath:=\alpha_1-\alpha_2.\label{vdef}
}
Obviously, \Math{-1\leq{}v\leq1}.
With these substitutions, the approximation of binomial distribution takes the form
\MathEqLab{P^*\vect{x,t}=
\e^{S^*}
	=\sqrt{\frac{2}{\pi{}\vect{1-v^2}t}}
	\exp\set{-\frac{1}{2t}\vect{\frac{x-v t}{\sqrt{1-v^2}}}^2}.}{Pappr}
The continuous variables \Math{x}, \Math{t}, and \Math{v} may be called, respectively, the ``space'', ``time'', and  ``velocity''.%
\footnote{In the paper \cite{Knuth}, which is devoted to the ``Zitterbewegung'' effect in the \Math{1+1} dimensional Dirac equation, a ``drift velocity'' is defined --- just like in \eqref{vdef} --- as the difference of probabilities of steps in opposite directions. 
It is shown that this definition leads to the relativistic velocity addition rule: \Math{w=\vect{u+v}/\vect{1+uv}}.
}
Expression \eqref{Pappr}  is the fundamental solution of the equation
\MathEqLab{
	\frac{\partial{}P^*\vect{x,t}}{\partial{}t}
	+v\frac{\partial{}P^*\vect{x,t}}{\partial{}x} 
	=\frac{\vect{1-v^2}}{2}
	\frac{\partial^2P^*\vect{x,t}}{\partial{}x^2}.}{heateq}
This equation  is called --- depending on interpretation of the function \Math{P^*\vect{x,t}} --- the \emph{heat}, or \emph{diffusion}, or \emph{Fokker-Plank} equation.
In the ``limit of the speed of light'' \Math{\cabs{v}=1} equation \eqref{heateq} turns into the \emph{wave equation}
\MathEq{\frac{\partial{}P^*\vect{x,t}}{\partial{}t}\pm\frac{\partial{}P^*\vect{x,t}}{\partial{}x}=0.}
Let us introduce the change of variables: \Math{t=T_H+t'} and \Math{x=vT_H+x'}.
If we assume that \Math{t'\ll{}T_H} (i.e., \Math{T_H} can be thought as a ``Hubble time'', and \Math{t'} as a ``typical time of observation''), then \eqref{Pappr} can be rewritten as
\MathEq{P^*\vect{x,t}=\sqrt{\frac{2}{\pi{}\vect{1-v^2}T_H}}\exp\set{-\frac{1}{2T_H}\vect{{\frac{x'-vt'}{\sqrt{1-v^2}}}}^2}+O\vect{\frac{t'}{T_H}}.}
The principal part of this expression is ``relativistically invariant''.
\subsection{The least action principle as the principle of selection of the most likely configurations}\label{leastaction}
Let us compare the exact probability distributions with their continuum approximations \emph{within} individual equivalence classes of relation \eqref{symmwalk}.
\paragraph{Exact distributions.}
In the binomial case, an equivalence class of \eqref{symmwalk} is defined by fixing the difference \Math{k_1-k_2=:X}.
We denote the equivalence class of sequences connecting the space-time points \Math{\vect{0,0}} and \Math{\vect{X,T}} by \Math{\lambda_{X,T}}.
The size of \Math{\lambda_{X,T}} is equal to \MathEq{N_{X,T}=\frac{T!}{k_1!k_2!}\equiv\frac{T!}{\vect{\frac{T+X}{2}}!\vect{\frac{T-X}{2}}!}~.}
The binomial distribution in terms of the variables \Math{X} and \Math{T} and parameter \Math{v} takes the form
\MathEq{P\vect{X,T}=\frac{T!}{\vect{\frac{T+X}{2}}!\vect{\frac{T-X}{2}}!}\vect{\frac{1+v}{2}}^{\frac{T+X}{2}}\vect{\frac{1-v}{2}}^{\frac{T-X}{2}}.}
Consider an increasing sequence of time instants (``times of observations'')
\MathEqLab{\tau=\set{T_0=0,\ldots,T_{i-1},T_{i},\ldots,T_{n}=T},~~T_{i-1}<{}T_{i}.}{times}
Let us select trajectories that pass through the sequence of spatial points
\MathEqLab{\chi=\set{X_0=0,\ldots,X_{i-1},X_{i},\ldots,X_{n}=X}}{Xpoints}
corresponding to sequence of times \eqref{times}.
Admissible trajectories must satisfy the inequality \Math{\cabs{X_i-X_{i-1}}\leq{}T_i-T_{i-1}} --- ``the light cone restriction''.
According to the \emph{conditional probability} rule, the probability of the trajectory \Math{\vect{\chi,\tau}} is equal to
\MathEqArrLab{P_{\chi,\tau}&\cmath=\frac{1}{P\vect{X,T}}\prod_{i=1}^nP\vect{X_i-X_{i-1},T_i-T_{i-1}}\nonumber\\
&\cmath=\frac{\vect{\frac{T+X}{2}}!\vect{\frac{T-X}{2}}!}{T!}\prod_{i=1}^n\frac{\vect{T_i-T_{i-1}}!}{\vect{\frac{T_i+X_i}{2}-\frac{T_{i-1}+X_{i-1}}{2}}!\vect{\frac{T_i-X_i}{2}-\frac{T_{i-1}-X_{i-1}}{2}}!}\enspace.\nonumber}
For a given sequence of time instants \eqref{times} one can formulate the problem of finding trajectories with maximum probability \Math{P_{\tau}=\max\limits_{\chi}P_{\chi,\tau}}. 
This can be done by searching among all admissible sequences \eqref{Xpoints}.
In the general case, there are many distinct trajectories with the same maximum probability,
i.e., we do not have here a ``deterministic'' trajectory.
\paragraph{Continuum approximation.}
To apply our reasoning to approximate distribution \eqref{Pappr},
we will consider the time sequence
\MathEq{\tau=\set{t_0=t_a,\ldots,t_{i-1},t_i,\ldots,t_n=t_b}}
together with respective sequences of spacial points
\MathEqLab{\chi=\set{x_0=x_a,\ldots,x_{i-1},x_{i},\ldots,x_{n}=x_b}.}{xpoints}
We assume that the time points are equidistant: \Math{t_i-t_{i-1}=\Delta{}t}, and we will use the notation \Math{\Delta{x_i}=x_i-x_{i-1}}.
\par
\noindent
Now the approximate probability of a trajectory \Math{\vect{\chi,\tau}} that connects the space-time points \Math{\vect{x_a,t_a}} and \Math{\vect{x_b,t_b}} takes the form
\MathEqArr{P^*_{\chi,\tau}=\sqrt{\frac{2}{\pi\vect{1-v^2}\vect{t_b-t_a}}}
&\cmath\exp\set{-\frac{1}{2\vect{t_b-t_a}}\vect{\frac{x_b-x_a-v\vect{t_b-t_a}}{\sqrt{1-v^2}}}^2}\\[4pt]
&\cmath\times\frac{Q_{\chi,\tau}}{A^n}\enspace,}
where
\MathEq{A=\sqrt{\frac{\pi\vect{1-v^2}\Delta{t}}{2}}}
and
\MathEqArrLab{Q_{\chi,\tau}&\cmath=\prod\limits_{i=1}^n\exp\set{-\frac{1}{2\Delta{t}}\vect{\frac{\Delta{x_i}-v\Delta{t}}{\sqrt{1-v^2}}}^2}\label{prodq}\\[2pt]
&\cmath=\exp\set{-\frac{1}{2}\sum\limits_{i=1}^n\vect{\frac{\Delta{x_i}/\Delta{t}-v}{\sqrt{1-v^2}}}^2\Delta{t}}.\label{sumprodq}}
The summation of factors in \eqref{prodq} over all values \Math{\Delta{x_i}\in\vect{-\infty,\infty}} reproduces correct normalization of probabilities
for any time slice: 
\MathEq{\int\limits_{-\infty}^\infty\exp\set{-\frac{1}{2\Delta{t}}\vect{\frac{\Delta{x_i}-v\Delta{t}}{\sqrt{1-v^2}}}^2}\frac{d\vect{\Delta{x_i}}}{2}=\sqrt{\frac{\pi\vect{1-v^2}\Delta{t}}{2}}\equiv{}A.}
Note that this normalization%
\footnote{A similar normalization is one of the cornerstones of the path integral formalism \cite{Feynman}.} 
is an approximation which is incompatible with the ``speed of light limitation'': \Math{-\Delta{t}\leq\Delta{x_i}\leq\Delta{t}.}
\par
Replacing the sequence of spacial points \eqref{xpoints} by a differentiable function \Math{x\vect{t}} such that \Math{x\vect{t_a}=x_a,~x\vect{t_b}=x_b},
introducing approximation \Math{\Delta{x_i}\approx{}\dot{x}\vect{t}\Delta{t}} and taking the limit \Math{n\rightarrow\infty} we can write instead of \eqref{sumprodq} the formula
\MathEq{\displaystyle{}Q_{\chi,\tau}\approx\exp\set{-\frac{1}{2}S\ordset{x\vect{t}}},}
where
\MathEq{S\ordset{x\vect{t}}=\int\limits_{t_a}^{t_b}Ldt=\int\limits_{t_a}^{t_b}\vect{\frac{\dot{x}\vect{t}-v}{\sqrt{1-v^2}}}^2dt\enspace.}  
Assuming for a while that \Math{v} 
depends on  \Math{t} and \Math{x}, we obtain the following Euler-Lagrange equation 
\MathEq{\frac{d}{dt}\frac{\partial{L}}{\partial{\dot{x}}}-\frac{\partial{L}}{\partial{x}}=0~\Rightarrow~
\ddot{x}\vect{1-v^2}+\dot{x}^2v\frac{\partial{v}}{\partial{x}}+2\dot{x}v\frac{\partial{v}}{\partial{t}}-v\frac{\partial{v}}{\partial{x}}-\vect{1+v^2}\frac{\partial{v}}{\partial{t}}=0.}
Clearly, this equation describes ``deterministic'' trajectories.
If we return to the initial assumption that \Math{v} does not depend on the space-time variables, then the Euler-Lagrange equation reduces to the form
\MathEq{\ddot{x}\vect{t}=0.}
This equation together with the boundary conditions gives the following formula for the extremals
\MathEq{x\vect{t}=\frac{x_b-x_a}{t_b-t_a}t+\frac{x_at_b-x_bt_a}{t_b-t_a},}
 i.e.,  the most probable trajectories are straight lines.
\section{Combinatorial models of quantum systems}\label{quant}
To build models that can reproduce quantum behavior, it is necessary to formulate the basic ingredients of quantum theory in a constructive way (for a more detailed consideration see \cite{KornyakPEPAN}).
\subsection{Constructive core of quantum mechanics}\label{quantcore}
In traditional matrix formulation quantum evolutions are described by unitary operators in a Hilbert space \Math{\Hspace}.
Evolution operators \Math{U} belong to a \alert{unitary representation} of the \emph{continuous} group \Math{\Aut{\Hspace}} of automorphisms of \Math{\Hspace}.
To make the problem constructive we should replace the group \Math{\Aut{\Hspace}} by some \emph{finite} group \Math{\wG} which should be empirically equivalent to (a subgroup of) \Math{\Aut{\Hspace}}.
\par
The theory of quantum computing \cite{Nielsen} proves the existence of \emph{finite} sets of universal \emph{quantum gates} that can be combined into unitary matrices which approximate to arbitrary precision any unitary operator. 
In other words, there exists a \emph{finitely generated} group \Math{\wG_{\infty}} which is a \emph{countable} dense subgroup of the continuous group \Math{\Aut{\Hspace}}.
\par
A group \Math{G} is called \emph{residually finite} \cite{Magnus}, if for every \Math{g\in{}G}, \Math{g\neq\id}, there exists a homomorphism \Math{\phi} from \Math{G} onto a finite group \Math{H}, such that \Math{\phi\vect{g}\neq\id}.
This means that any relation between the elements of \Math{G} can be modeled by a relation between the elements of a finite group.
This can be illustrated by analogy with the widely used in physics technique, when an infinite space is replaced by, for example, a torus 
whose size is sufficient to hold the data related to a particular problem.
\par
According to the theorem of A.I. Mal'cev \cite{Malcev}, every finitely generated group of matrices over any field is residually finite.
Thus we have the sequence of transitions from the group with cardinality of \emph{continuum} through a \emph{countable} group to a \emph{finite} group:
\Math{{\Aut{\Hspace}\xrightarrow{\text{approximation}}\wG_{\infty}\xrightarrow{\text{homomorphism}}\wG}\,.}
\subsection{Permutations and natural quantum amplitudes}\label{quantperm}
As is well known, any linear representation of a finite group is unitary. 
Any representation of a finite group is a subrepresentation of some permutation representation (see, e.g., \cite{Hall,Serre,Wielandt,Cameron,Dixon}).
Let \Math{\repq} be a representation of \Math{\wG} in a \Math{\adimH}-dimensional Hilbert space \Math{\Hspace_{\adimH}}.
Then \Math{\repq} can be embedded into a permutation representation \Math{\regrep} of \Math{\wG} in an \Math{\wSN}-dimensional Hilbert space \Math{\Hspace_{\wSN}}, where \Math{\wSN\geq\adimH}. 
The representation \Math{\regrep} is equivalent to an action of \Math{\wG} on a set of things \Math{\wS=\set{\ws_1,\ldots,\ws_\wSN}} by permutations.
In the proper case \Math{\wSN>\adimH}, the embedding has the structure
\MathEqLab{
\transmatr^{-1}\regrep\transmatr
=\Vtwo{
\left.
\begin{aligned}
\!\IrrRep{1}&\\[-2pt]
&\hspace{8pt}\mathrm{V}
\end{aligned}
\right\}\Hspace_{\wSN-\adimH}
}{
\left.
\hspace{29pt}
{\repq}
\right\}\Hspace_{\adimH}
},\hspace{10pt}
\Hspace_{\wSN} = \Hspace_{\wSN-\adimH}\oplus\Hspace_{\adimH},
}{embed}	
where \Math{\IrrRep{1}} is the trivial one-dimensional representation, mandatory for any permutation representation; \Math{\mathrm{V}} is a subrepresentation,  which may be missing. \Math{\transmatr} is a matrix of transition from the basis of the  representation \Math{\regrep} to the basis in which the permutation space \Math{\Hspace_{\wSN}} is split into the invariant subspaces \Math{\Hspace_{\wSN-\adimH}} and \Math{\Hspace_{\adimH}}. 
Evolutions in the spaces  \Math{\Hspace_{\adimH}} and \Math{\Hspace_{\wSN-\adimH}}
are \emph{independent} since both spaces are invariant subspaces of \Math{\Hspace_{\wSN}}. So we can treat the data in \Math{\Hspace_{\wSN-\adimH}} as ``hidden parameters'' with respect to the data in \Math{\Hspace_{\adimH}}.
\par
A trivial approach would be to set arbitrary (e.g., zero) data in the complementary subspace \Math{\Hspace_{\wSN-\adimH}}. 
This approach is not interesting since it is not falsifiable by means of standard quantum mechanics. 
In fact, it leads to standard quantum mechanics \emph{modulo} the empirically unobservable distinction between the ``finite'' and the ``infinite''. 
The only difference is technical: we can replace the linear algebra in the \Math{\adimH}-dimensional space \Math{\Hspace_{\adimH}} by permutations of \Math{\wSN} things.
\par
A more promising approach requires some changes in the concept of quantum amplitudes. 
We assume \cite{KornyakPEPAN,Kornyak12,Kornyak13a} that quantum amplitudes are projections onto invariant subspaces of vectors of multiplicities of elements of the set \Math{\wS} on which the group \Math{\wG} acts by permutations.
The vectors of multiplicities 
\MathEqLab{\barket{n} = \Vthree{n_1}{\vdots}{n_{\wSN}}}{multvect} 
are elements of the \emph{module} \Math{\natmod_\wSN = \N^\wSN}, where \Math{\N=\set{0,1,2,\ldots}} is the semiring of natural numbers.
Initially we deal with the natural permutation representation of \Math{\wG} in the module
\Math{\natmod_\wSN}.
Using  the fact that all eigenvalues of any linear representation of a finite group are roots of unity, we can turn the module \Math{\natmod_\wSN} into a Hilbert space \Math{\Hspace_{\wSN}}.
It is sufficient to add \Math{\period}th roots of unity to the natural numbers to form a \emph{semiring}, which we denote by \Math{\N_\period}. 
The natural number \Math{\period}, called conductor, is (a divisor of) the \emph{exponent} of \Math{\wG}, which is defined as the least common multiple of the orders of elements of \Math{\wG}.
In the case  \Math{\period\geq2} the \emph{negative integers} can be introduced and the semiring \Math{\N_\period} becomes a \emph{ring of cyclotomic integers}. To complete the conversion of the 
module \Math{\natmod_\wSN} into the Hilbert space \Math{\Hspace_{\wSN}}, we introduce the \emph{cyclotomic field}
\Math{\Q_\period} as a field of fractions of the ring \Math{\N_\period}.%
\footnote{By taking into account symmetries of a specific problem, we can use instead of \Math{\Q_\period} some its subfield, 
an \emph{abelian number field} \Math{\NF\leq\Q_\period}.}
If \Math{\period\geq3}, then \Math{\Q_\period} is a dense subfield of the field of complex numbers \Math{\C}.  
In fact,  algebraic properties of elements of \Math{\Q_\period} are quite sufficient for all our purposes --- for example, complex conjugation corresponds to the transformation \Math{\runi{}^k\rightarrow\runi{}^{\period-k}} for roots of unity  --- so we can forget the possibility to embed \Math{\Q_\period} into \Math{\C} (as well as the very existence of the field \Math{\C}).
\par
Thus, we will assume that \Math{\Hspace_{\adimH}} in decomposition \eqref{embed} is a Hilbert space over the field \Math{\Q_\period}, and quantum amplitudes are elements of \Math{\Hspace_{\adimH}} of the form \Math{\barket{\psi}=\projector{\repq}\barket{n}},
where \Math{\projector{\repq}} is the projection operator from the module \Math{\natmod_\wSN} onto the space \Math{\Hspace_{\adimH}} corresponding to the subrepresentation \Math{\repq} in \eqref{embed}.
\subsection{Measurements and the Born rule}\label{Born}
The general scheme of measurements%
\footnote{To avoid inessential technical complications we consider here only the case of pure states.}
in quantum mechanics is reduced to the following.
\begin{itemize}
	\item
A partition of the Hilbert space into mutually orthogonal subspaces is given:
\MathEq{\Hspace=\Hspace_1\oplus\cdots\oplus\Hspace_i\oplus\cdots.}
Typically \Math{\Hspace_i} are eigenspaces of some Hermitian operator \Math{A} (i.e. \Math{A=A^\dagger}) called an ``observable''.
	\item
There is a measuring device configured to select a state \Math{\barket{\phi}\in\Hspace_i} of a quantum system.
	\item
A result of a single measurement
\Math{=
\begin{cases}
\mathrm{Yes},& \text{\color{black}measuring device responds;}\\
\mathrm{No},& \text{\color{black}no response.}
\end{cases}
}
\par
In accordance with the \emph{projection postulate}, the output \Math{\mathrm{Yes}}  is interpreted as transition of the system into the state \Math{\barket{\phi}} \emph{after} the measurement.
\par
If \Math{\Hspace_i} is an eigenspace of an observable \Math{A} with eigenvalue \Math{a}, it is said that the ``outcome of the measurement is equal to''   \Math{a}.
	\item
Relative number of \Math{\mathrm{Yes}} in a set of measurements is described by the Born formula.
\end{itemize}
The Born rule%
\footnote{There have been many attempts to derive the Born rule from the other physical assumptions --- the Schr\"{o}dinger equation, many-worlds interpretation, etc. However, Gleason's theorem \cite{Gleason} shows that the Born rule is a logical consequence of the very definition of a Hilbert space and has nothing to do with the laws of evolution of physical systems.}
states that the probability to register a particle described by the amplitude \Math{\barket{\psi}} by an apparatus configured to select the amplitude \Math{\barket{\phi}} is
	\MathEq{\ProbBorn{\phi}{\psi} = \frac{\cabs{\inner{\phi}{\psi}}^2}
		{\inner{\phi}{\phi}\inner{\psi}{\psi}}~.}
In the ``finite'' background the only reasonable interpretation of probability is the \emph{frequency interpretation}: 
probability is the ratio of the number of ``favorable'' combinations to the total number of combinations. So we expect that \Math{\ProbBorn{\phi}{\psi}} must be a \emph{rational number} if everything is arranged correctly.
Thus, in our approach the usual \alert{non-constructive} contraposition
--- \alert{complex numbers} as intermediate values against \alert{real numbers} as observable values 
--- is replaced by the \alert{constructive} one --- \alert{irrationalities} against \alert{rationals}.
From the constructive point of view, there is no fundamental difference between irrationalities and constructive complex	numbers: both are elements of algebraic extensions.
\subsection{Illustration: natural amplitudes and invariant subspaces of permutation representation}
\begin{figure}[!hB]
\centering
\includegraphics[width=0.5\textwidth]{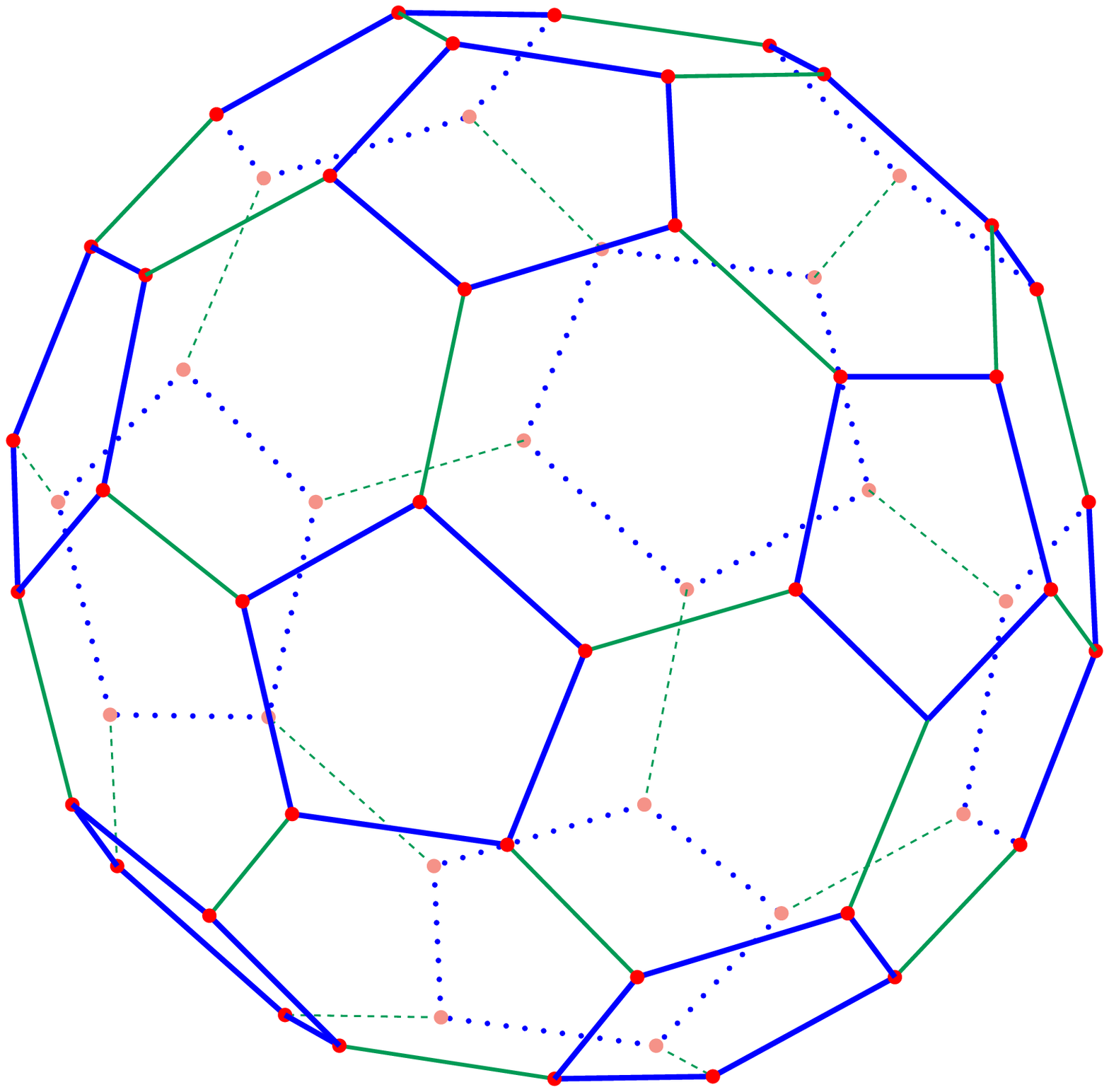}
\caption{Cayley graph of \Math{\AltG{5}} for presentation \eqref{Enpresent}. 
Pentagons, hexagons and links between adjacent pentagons correspond to the relators \Math{a^5}, \Math{\vect{a{\ccom{}b}}^3} and  \Math{{\ccom{}b}^2}, respectively.}
\label{Enbucky}
\end{figure}

Consider the action of the alternating group  \Math{\AltG{5}} on the vertices of the icosahedron.
The group  \Math{\AltG{5}} has a \emph{presentation} of the form 
\MathEqLab{\AltG{5} = \left\langle{}a, {\ccom{}b}\left|\,a^5={\ccom{}b}^2=(a{\ccom{}b})^3=\id\right.\right\rangle.}{Enpresent}
The Cayley graph of this presentation is shown in Figure \ref{Enbucky}.
\par\noindent
\Math{\AltG{5}} has five irreducible representations: the trivial \Math{\IrrRep{1}} and four faithful representations \Math{\IrrRep{3}, \IrrRep{3'}, \IrrRep{4}, \IrrRep{5}};
and three \emph{primitive}%
\footnote{A transitive action of a group on a set is called \emph{primitive} \cite{Wielandt}, if there is no \emph{non-trivial partition} of the set, invariant under the action of the group.}
permutation representations having the following decompositions into the irreducible components:
\Math{\PermRep{5}\cong\IrrRep{1}\oplus\IrrRep{4}},~~
 \Math{\PermRep{6}\cong\IrrRep{1}\oplus\IrrRep{5}},~~and
 \Math{\PermRep{10}\cong\IrrRep{1}\oplus\IrrRep{4}\oplus\IrrRep{5}.}
\par
\begin{figure}[!h]
\centering
\includegraphics[width=0.5\textwidth]{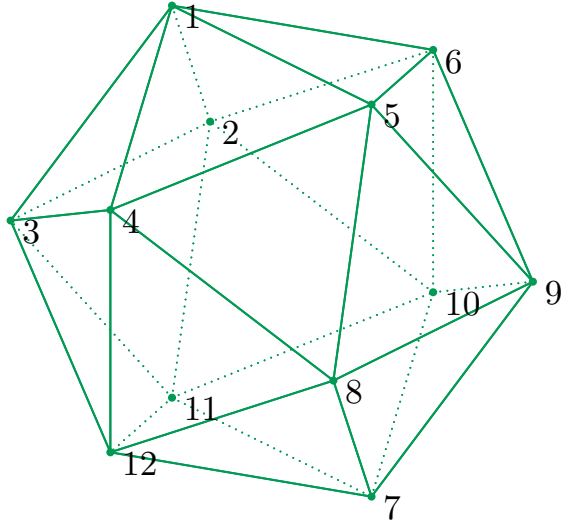}
\caption{Icosahedron. Invariant blocks are pairs of opposite vertices.}
\label{Enico}
\end{figure}

The action of \Math{\AltG{5}} on the icosahedron vertices \Math{\wS=\set{1,\ldots,12}} is transitive, but \emph{imprimitive} with the non-trivial partition into the following blocks 
\MathEq{\set{\mid{}B_1\mid\cdots\mid{}B_i\mid\cdots\mid{}B_6\mid}\equiv
\set{\mid1,7\mid\cdots\mid{}i,i+6\mid\cdots\mid6,12\mid},}
assuming the vertex numbering shown in Figure \ref{Enico}.
Each block \Math{B_i} consists of a pair of opposite vertices of the icosahedron.
Permutation representation of the action of \Math{\AltG{5}} on the icosahedron vertices has the following decomposition into irreducible components
\MathEqLab{\transmatr^{-1}\vect{\PermRep{12}}\transmatr
	=\IrrRep{1}\oplus\IrrRep{3}\oplus\IrrRep{3'}\oplus\IrrRep{5},}{decoico}
where \Math{\transmatr} is a matrix of transition from the ``permutation'' to the ``splitting'' basis.
\par
Actually there is no necessity to compute transformation matrices like \Math{\transmatr} in \eqref{decoico} explicitly.
There is a way \cite{KornyakPEPAN} to express invariant scalar products in invariant subspaces in terms of easily computable matrices of \emph{orbitals} \cite{Cameron,Dixon}, 
i.e., orbits of the action  of a group \Math{\wG} on the Cartesian product \Math{\wS\times\wS}.
\par
For the action of \Math{\AltG{5}} on the set of icosahedron vertices, the matrices of orbitals have the form
\MathEqLab{\baseform{1}=\idmat_{12},~ \baseform{2}=\Mtwo{0}{\idmat_{6}}{\idmat_{6}}{0},~\baseform{3}=\Mtwo{X}{Y}{Y}{X},~\baseform{4}=\Mtwo{Y}{X}{X}{Y},}{orbitals}
where \Math{\idmat_n} is \Math{n\times{}n} identity matrix, and
\MathEq{
X=
\bmat
0&1&1&1&1&1\\
1&0&1&0&0&1\\
1&1&0&1&0&0\\
1&0&1&0&1&0\\
1&0&0&1&0&1\\
1&1&0&0&1&0
\emat,~
Y=
\bmat
0&0&0&0&0&0\\
0&0&0&1&1&0\\
0&0&0&0&1&1\\
0&1&0&0&0&1\\
0&1&1&0&0&0\\
0&0&1&1&0&0
\emat.
}
In terms of matrices \eqref{orbitals} the invariant bilinear forms (scalar products) corresponding to decomposition \eqref{decoico} take the form
\begin{align*}\cmath{}
\bornform{\IrrRep{1}} =&\cmath{} \frac{1}{12}\vect{\baseform{1}+\baseform{2}+\baseform{3}+\baseform{4}},
\\\cmath{}
\bornform{\IrrRep{3}} =&\cmath{} \frac{1}{4}\vect{\baseform{1}-\baseform{2}-\frac{1+2\runi{}^2+2\runi{}^3}{5}\baseform{3}+\frac{1+2\runi{}^2+2\runi{}^3}{5}\baseform{4}},
\\\cmath{}
\bornform{\IrrRep{3'}} =&\cmath{} \frac{1}{4}\vect{\baseform{1}-\baseform{2}+\frac{1+2\runi{}^2+2\runi{}^3}{5}\baseform{3}-\frac{1+2\runi{}^2+2\runi{}^3}{5}\baseform{4}},
\\\cmath{}
\bornform{\IrrRep{5}} =&\cmath{} \frac{5}{12}\vect{\baseform{1}+\baseform{2}-\frac{1}{5}\baseform{3}-\frac{1}{5}\baseform{4}},
\end{align*}
where \Math{\runi{}} is a 5th primitive root of unity.
It is easy to verify that the cyclotomic integer \Math{1+2\runi{}^2+2\runi{}^3} is equal to  \Math{-\sqrt{5}}.
\par
Let us consider the scalar products of projections of ``natural'' vectors.
If projections of vectors with natural components \Math{{m}=\vect{m_1,\ldots,m_{12}}^T} and \Math{n=\vect{n_1,\ldots,n_{12}}^T} onto the invariant subspaces corresponding to \Math{\alpha=\IrrRep{1}, \IrrRep{3}, \IrrRep{3'}, \IrrRep{5}} are \Math{\Phi_\alpha} and \Math{\Psi_\alpha}, respectively, then \Math{\inner{\Phi_\alpha\!}{\!\Psi_{\alpha}}= \innerform{m}{\bornform{\alpha}}{n}}. That is, we have
\begin{align}\cmath{}
\inner{\Phi_{\IrrRep{1}}}{\Psi_{\IrrRep{1}}}&\cmath{}=\frac{1}{12}\Bigl(\innerform{m\!}{\baseform{1}}{\!n}\!+\!\innerform{m\!}{\baseform{2}}{\!n}\!+\!\innerform{m\!}{\baseform{3}}{\!n}\!+\!\innerform{m\!}{\baseform{4}}{\!n}\Bigr),\label{prod1}
\\\cmath{}
\inner{\Phi_{\IrrRep{3}}}{\Psi_{\IrrRep{3}}}&\cmath{}=\frac{1}{4}\vect{\innerform{m\!}{\baseform{1}}{\!n}\!-\!\innerform{m\!}{\baseform{2}}{\!n}\!+\!\frac{\sqrt{5}}{5}\Bigl(\innerform{m\!}{\baseform{3}}{\!n}\!-\!\innerform{m\!}{\baseform{4}}{\!n}\Bigr)},\nonumber
\\\cmath{}
\inner{\Phi_{\IrrRep{3'}}}{\Psi_{\IrrRep{3'}}\!}&\cmath{}=\frac{1}{4}\vect{\innerform{m\!}{\baseform{1}}{\!n}\!-\!\innerform{m\!}{\baseform{2}}{\!n}\!-\!\frac{\sqrt{5}}{5}\Bigl(\innerform{m\!}{\baseform{3}}{\!n}\!+\!\innerform{m\!}{\baseform{4}}{\!n}\Bigr)},\nonumber
\\\cmath{}
\inner{\Phi_{\IrrRep{5}}}{\Psi_{\IrrRep{5}}}&\cmath{}=\frac{5}{12}\vect{\innerform{m\!}{\baseform{1}}{\!n}\!+\!\innerform{m\!}{\baseform{2}}{\!n}\!-\!\frac{1}{5}\Bigl(\innerform{m\!}{\baseform{3}}{\!n}\!+\!\innerform{m\!}{\baseform{4}}{\!n}\Bigr)}.\nonumber
\end{align}
Let us give two remarks on these expressions:
\begin{itemize}
	\item
Scalar product \eqref{prod1} can be written as 
\MathEq{\inner{\Phi_{\IrrRep{1}}}{\Psi_{\IrrRep{1}}}=\frac{1}{12}\vect{m_1+m_2+\cdots+m_{12}}\vect{n_1+n_2+\cdots+n_{12}}.}
This is the general case: any permutation representation of any group contains the trivial one-dimensional subrepresentation with the scalar product like \eqref{prod1}: 
\MathEq{\inner{\Phi_{\IrrRep{1}}}{\Psi_{\IrrRep{1}}}=\frac{1}{\wSN}\vect{\sum_{i=1}^{\wSN}{m_i}}\vect{\sum_{i=1}^{\wSN}{n_i}}.}
The trivial subrepresentation can be interpreted as the ``counter of particles'', since the linear permutation invariant \Math{\sum_{i=1}^{\wSN}{n_i}} is the total number of elements from \Math{\wS} in the ensemble described by the vector \Math{n}.
	\item 
The Born probabilities for subrepresentations \Math{\IrrRep{3}} and \Math{\IrrRep{3'}} contain irrationalities that contradicts the frequency interpretation of probability for finite sets.
Obviously, this is a consequence of the imprimitivity: one can not move an icosahedron vertex without simultaneous movement of its opposite.
To resolve the contradiction, mutually conjugate subrepresentations  \Math{\IrrRep{3}} and \Math{\IrrRep{3'}} must be considered together.
The scalar product 
\MathEq{\inner{\Phi_{\IrrRep{3\oplus3'}}}{\Psi_{\IrrRep{3\oplus3'}}}=\frac{1}{2}\Bigl(\innerform{m}{\baseform{1}}{n}-\innerform{m}{\baseform{2}}{n}\Bigr)} 
in the six-dimensional subrepresentation \Math{\IrrRep{3}\oplus\IrrRep{3'}} always gives rational Born's probabilities for vectors of multiplicities defined as in \eqref{multvect}.	
\end{itemize}
\subsection{Quantum evolution}\label{quantumevolution}
In standard quantum mechanics an elementary step of evolution is described by the Schr\"{o}dinger equation
\MathEq{i\frac{\mathrm{d}}{\mathrm{d}t}\barket{\psi}=H\barket{\psi}.}
In quantum mechanics based on a finite group \Math{\wG} a step of evolution has the form  
\MathEq{\barket{\psi_{t+1}}=U\barket{\psi_t},}
where \Math{U=\repq\vect{g}},  \Math{g\in\wG} and \Math{\repq} is an unitary representation of \Math{\wG}.
In this case, there is no need for a Hamiltonian, though, for comparison purposes,  it can be introduced by the formula
\MathEq{H=i\ln{}U 
=\pi\!\vect{\alpha_0\idmat+\alpha_1U+\cdots+\alpha_{n-1}U^{n-1}},}
where \Math{\idmat} is the unit matrix; \Math{n} is the period of \Math{U}, i.e. \Math{U^n=\idmat}; \Math{\alpha_k\in\NF} are easily computable coefficients; \Math{0\leq{k}<n}.
The energy levels (eigenvalues) of \Math{H} are \Math{\displaystyle{}E_k=\frac{2\pi{k}}{n}}.
The non-algebraic (transcendental) number \Math{\pi} appears here as the result of summation of infinite series --- the natural logarithm is essentially an infinite construct.
\par
Note that a single unitary evolution is physically trivial, as it describes only a change of coordinates (``rotation'') in a Hilbert space.
Namely, for the evolution of a pair of vectors \Math{\barket{\phi_T}=U\barket{\phi_0},~\barket{\psi_T}=U\barket{\psi_0}}, we have
\MathEq{\inner{\phi_T\!}{\!\psi_T}=\innerform{\phi_0}{U^{\dagger}U}{\psi_0}\equiv\inner{\phi_0\!}{\!\psi_0}.}
This means that a single deterministic evolution can not provide physically observable effects.
Thus, a collection of different evolutions is needed. 
Suppose that the operators of evolution belong to a unitary representation \Math{\repq} of a group \Math{\wG}.
Then two different evolutions \Math{U} and \Math{V} can be represented as \Math{U=\repq\vect{g_1g_2\cdots{}g_T}} and \Math{V=\repq\vect{f_1f_2\cdots{}f_T}}, where \Math{g_1,\cdots,g_T; f_1,\cdots,f_T\in\wG}.
These evolutions provide a nontrivial physical effect  if 
\MathEq{\inner{\phi_T\!}{\!\psi_T}=\innerform{\phi_0}{V^{\dagger}U}{\psi_0}\neq\inner{\phi_0\!}{\!\psi_0}\Longleftrightarrow{V^{\dagger}U}=\repq\vect{f_T^{-1}\cdots{}f_1^{-1}g_1\cdots{}g_T}\neq\idmat}
or
\MathEq{h=f_T^{-1}\cdots{}f_1^{-1}g_1\cdots{}g_T\neq\id,}
where \Math{\id} is the identity of \Math{\wG}. 
The expression \Math{h=f_T^{-1}\cdots{}f_1^{-1}g_1\cdots{}g_T} is called the \emph{holonomy} at the point \Math{T} of principal \Math{\wG}-connection.
In differential geometry, infinitesimal analogue of holonomy is called the \emph{curvature} of the corresponding connection.
As is well known, all fundamental physical forces are represented in the gauge theories as curvatures of appropriate connections.
\par
Observations support the view that fundamental indeterminism is really the \textit{modus operandi} of nature.%
\footnote{
There are persistent attempts to develop a deterministic version of quantum mechanics, as if determinism were  a ``synthetic \textit{a priori} judgment'' 
--- an inevitable (though not deducible from logic alone) necessity.
However, since the time of Kant up to now there are no convincing  evidences of the very existence of judgments of this kind.
More likely, the belief in determinism is a mental habit formed by long macroscopic experience.
}
In our view this indeterminism arises from a fundamental impossibility to trace the identity of indistinguishable objects during their evolution.
Hermann Weyl discussed this issue in detail in \cite{Weyl}. 
More formally, identification of objects at different time points is provided by a {connection} ({parallel transport}).
\par
In our setting we consider dynamical system as a fiber bundle  \Math{\vect{\FBT,\Time,\wS,\wG,\tau}} over discrete time \Math{\Time=\ordset{0,1,\ldots,T}}, 
where \emph{typical fiber} \Math{\wS=\set{\ws_1,\ldots,\ws_{\wSN}}} is \alert{canonical set} of states,
\emph{structural group} 	
\MathEqLab{\wG=\set{\wg_1,\ldots,\wg_m,\ldots,\wg_{\wGN}}\leq\Perm{\wS}\cong\SymG{\wSN}}{defG} 
is the \alert{group of symmetries} of states,
\Math{\tau} is a projection \Math{\FBT\rightarrow{}\Time}.
\par
\alert{Connection} (\emph{parallel transport}) \Math{\mbox{\raisebox{3pt}{\Math{\Partransportt}}}_{t_1t_2}\in\wG} defines isomorphism between the fibers at different times of observations: 
\MathEq{\wS_{t_2}=\wS_{t_1}\mbox{\raisebox{3pt}{\Math{\Partransportt}}}_{t_1t_2}\,.}
There is no objective way to choose  the ``correct'' value for the connection. 
\textit{A priori}, any element of \Math{\wG} may serve as  \Math{\mbox{\raisebox{3pt}{\Math{\Partransportt}}}_{t_1t_2}}.
In continuous gauge theories, the gauge fields (fields of connections) are determined from the principle of least action using Lagrangians chosen for different reasons.
For example, in the case of the Yang-Mills theory (covering also the case of Maxwell equations), the Lagrangian \Math{L_{\mathrm{YM}}=\mathrm{Tr}\left[F\wedge\star{}F\right]} is used,
where \Math{F} is the curvature form of a gauge connection, \Math{\star} denotes the \emph{Hodge conjugation}.
Analysis of the structure of \Math{L_{\mathrm{YM}}} in the discrete approximation \cite{Oeckl} shows that it can be expressed in terms of traces of the fundamental representation  of holonomies of a gauge group.
\subsection{Combinatorial models of gauge and quantum evolution}\label{combmod}
Consider a simple combinatorial model involving random choice of the rules for identification of states of dynamical systems at different points of time.
The time of our model is the sequence \Math{\Time = \ordset{0,1,\ldots,t-1,t,\ldots,T}}.
A parallel transport connecting the initial and the final time points can be decomposed into the product of elementary steps: 
\MathEqLab{\mbox{\raisebox{3pt}{\Math{\Partransportt}}}_{0,1}\cdots\mbox{\raisebox{3pt}{\Math{\Partransportt}}}_{t-1,t}\cdots\mbox{\raisebox{3pt}{\Math{\Partransportt}}}_{T-1,T}\,.}{connect}
We assume that any elementary step is an element of group \eqref{defG} with probability independent of time: 
\MathEq{\Prob\vect{\mbox{\raisebox{3pt}{\Math{\Partransportt}}}_{t-1,t}=\wg_m\in\wG}=\alpha_m,~~~\sum_{m=1}^\wGN\alpha_m=1\,.}
All possible paths \eqref{connect} form the set of microstates \Math{U}.
The microstate corresponding to \eqref{connect} is the sequence \Math{u={\wg_{m_1},\ldots,\wg_{m_t},\ldots,\wg_{m_T}}}, where \Math{\wg_{m_t}=\mbox{\raisebox{3pt}{\Math{\Partransportt}}}_{t-1,t}.} 
Its probability is the product \Math{p_u=\alpha_{m_1}\cdots\alpha_{m_t}\cdots\alpha_{m_T}}.
\par
We can define a natural equivalence relation on \Math{U} as the triviality of the holonomy of a pair of paths:
\MathEq{\mbox{\raisebox{3pt}{\Math{\Partransportt}}}_{0,1}\cdots\mbox{\raisebox{3pt}{\Math{\Partransportt}}}_{T-1,T}\sim\mbox{\raisebox{3pt}{\Math{\gamma}}}_{0,1}\cdots\mbox{\raisebox{3pt}{\Math{\gamma}}}_{T-1,T}\Longleftrightarrow
\mbox{\raisebox{3pt}{\Math{\gamma}}}_{T-1,T}^{-1}\cdots\mbox{\raisebox{3pt}{\Math{\gamma}}}_{0,1}^{-1}\mbox{\raisebox{3pt}{\Math{\Partransportt}}}_{0,1}\cdots\mbox{\raisebox{3pt}{\Math{\Partransportt}}}_{T-1,T}=\id\,.}
This equivalence allows us to define \Math{\wGN} macrostates \Math{\lambda_1,\ldots,\lambda_\wGN}.
Statistical evolution of all the macrostates can be calculated simultaneously by a simple algorithm.
The distribution of the macrostates at the moment \Math{T} is the following element of the group algebra
\MathEq{A_T\equiv{}P_{\lambda_1}\wg_1+P_{\lambda_2}\wg_2+\cdots+P_{\lambda_\wGN}\wg_\wGN=\vect{\alpha_1\wg_1+\alpha_2\wg_2+\cdots+\alpha_\wGN\wg_\wGN}^T.}
The algorithm of \emph{binary exponentiation} computes this expression by performing \Math{O\vect{\log{T}}} multiplications.
In simple cases the probabilities \Math{P_{\lambda_m}} can be written explicitly, e.g., for the cyclic group \Math{\wG=\CyclG{\wGN}} we have (assuming \Math{\wg_1=\id})
\MathEq{P_{\lambda_m}=\sum_{\substack{k_1+k_2+\cdots+k_{\wGN}=T\\
k_2+2k_3+\cdots+\vect{\wGN-1}k_{\wGN}\,\equiv\,{m}\pmod{\wGN}}}\frac{T!}{k_1!k_2!\cdots{k_{\wGN}!}}\alpha_1^{k_1}\alpha_2^{k_2}\cdots\alpha_{\wGN}^{k_{\wGN}}\,.} 
\par
Having a representation \Math{\repq} of the group \Math{\wG} in a Hilbert space \Math{\Hspace} we can associate with gauge evolution \eqref{connect} the quantum evolution:
\MathEq{\barket{\psi_T}=\repq\vect{\mbox{\raisebox{3pt}{\Math{\Partransportt}}}_{T-1,T}^{-1}\cdots\mbox{\raisebox{3pt}{\Math{\Partransportt}}}_{t-1,t}^{-1}\cdots\mbox{\raisebox{3pt}{\Math{\Partransportt}}}_{0,1}^{-1}}\barket{\psi_0},~~~\psi_0,\psi_T\in\Hspace\,.}
\par
Simulation of many important features of quantum behavior requires models involving spatial structures explicitly.
The set of states of a system with space is the set of functions
\MathEqLab{\wS=\lSX,}{statefunc}
where  
\MathEq{\X=\set{\x_1,\ldots, \x_\XN}}
is a \emph{space}, and
\MathEq{\lS=\set{\ls_1,\ldots,\ls_\lSN}}
is a set of \emph{local states}.
Having the groups of \emph{spatial}
\MathEq{\sG=\set{\sg_1,\ldots,\sg_\sGN}\leq\Perm{\X}}
and \emph{internal} 
\MathEq{\iG=\set{\ig_1,\ldots, \ig_\iGN}\leq\Perm{\lS}}
symmetries we can construct a symmetry group of the whole system .
This group, having a structure of the \emph{wreath product}  
\MathEq{\wG=\iG\wr_\X\sG\cong\iGX\rtimes\sG,}
acts on the set \Math{\wS} given by formula \eqref{statefunc}.
\par
It is worth to say a few words about the most common  quantum models with spatial structures: \emph{quantum cellular automata} \cite{McDonald} and \emph{quantum walks} \cite{QWalk}.
It is proved that models of both types are able to perform any quantum computation, i.e., they can simulate quantum Turing machines.
\par
The Hilbert space of a \textbf{quantum cellular automaton} has the form
\MathEq{\Hspace=\Hspace_{\lS}^{\otimes\X},} 
where \Math{\X} is usually a \Math{d}-dimensional lattice: \Math{\Z^d} or its finite counterpart \Math{\Z_N^d} (one can also take an arbitrary regular graph as a lattice \Math{\X}); 
\Math{\Hspace_{\lS}} is a Hilbert space associated with a set \Math{\lS} of local states of sites \Math{x\in\X}.
It is assumed that there is a local update rule \Math{U_x}, which is a unitary operator acting on the Hilbert space \Math{\Hspace_{\lS}^{\otimes\mathcal{N}_x}},
where \Math{\mathcal{N}_x} is a neighborhood of the point \Math{x}. 
Since the neighborhoods of different points may intersect, some effort should be made to ensure global unitary.
To provide the required compatibility several different definitions of quantum cellular automata were proposed.
For properly defined automaton the local updates can be combined into a unitary operator \Math{U} on \Math{\Hspace} that describes an elementary step of evolution of the whole system.
Then the evolution of the system is defined by the operator \Math{U^T}.
\par
The spatial structure in a model of \textbf{quantum walk} is a \Math{k}-regular graph \Math{\X}.
In the most usual case \Math{k=2}, the space  \Math{\X} is taken to be either \Math{\Z} or \Math{\Z_N}.
Let \Math{\Hspace_\X} be the Hilbert space spanned by the vertices of \Math{\X}.
The construction of a quantum walk uses also an auxiliary \Math{k}-dimensional Hilbert space \Math{\Hspace_C}, the ``coin space'',
and a fixed unitary ``coin operator'' \Math{C} acting on \Math{\Hspace_C}. 
A typical coin operator in the case \Math{k>2} is the \emph{Grover coin} (\emph{Grover's diffusion operator}): 
\MathEq{G=2\barket{\psi}\brabar{\psi}-\idmat_k,~~~\text{\color{black}where}~~ \barket{\psi}=\frac{1}{\sqrt{k}}\Vthree{1}{\vdots}{1}.} 
In the case of one-dimensional quantum walk (\Math{k=2}) many different one-qubit gates, like the \emph{Hadamard gate} etc., are used.
In particular, the Grover coin \Math{G} coincides with the \emph{Pauli-X gate} at \Math{k=2}.
The Hilbert space of the whole system is the product
\MathEq{\Hspace=\Hspace_\X\otimes\Hspace_C.}
Roughly speaking, the coin operator \Math{C} ``selects directions of spatial shifts''.
The spatial shifts are performed by an unitary \emph{shift operator} \Math{S} acting on \Math{\Hspace_\X} at conditions given by  the coin \Math{C}.
\Math{T} steps of evolution of the system are performed by the transformation \Math{U^T}, where \Math{U} is the following combination of the coin and shift operators
\MathEq{U=S\vect{\idmat_k\otimes{}C}.} 
\section{Summary}
Starting with the idea that any problem that has a meaningful empirical content can be formulated in constructive finite terms, we consider the possibility of derivation of many important elements of physical theories in the framework of discrete combinatorial models.
We show that such concepts as continuous symmetries, the principle of least action, Lagrangians, deterministic evolution equations can be obtained by applying the large number approximation to expressions for sizes of certain equivalence classes of combinatorial structures.
\par
We adhere to the view  that quantum behavior can be explained by the fundamental impossibility to trace identity of indistinguishable objects in the process of their evolution.
Gauge connection is that structure which provides the identity: that is why the gauge fields are so important in quantum theory.
\par
Using general mathematical arguments we show that any quantum problem can be reduced to permutations.
Quantum interferences are phenomena observed in invariant subspaces of permutation representations and expressed in terms of permutation invariants.
In particular, this approach gives an immediate explanation for the appearance of complex numbers and unitarity in the formalism of quantum theory.
\par
We consider some approaches to the construction of discrete models of quantum behavior and related models describing evolution of gauge connections.


\Acknowledgements{I am grateful to 
A.Yu.\,Blinkov, V.P.\,Gerdt, A.M.\,Ishkhanyan and S.I.\,Vinitsky for many insightful discussions and comments.}




\end{document}